\documentclass[nopreprint]{jasatex}
\usepackage{amssymb}
\usepackage{amsmath}
\usepackage{graphicx}
\usepackage[utf8]{inputenc}
\usepackage[T1]{fontenc}
\usepackage{setspace}
\usepackage{multirow}
\usepackage{booktabs}
\usepackage{url}
\newcommand{\MDversion}{arXiv}

\pdfoutput=1

\usepackage[hidelinks,
pdftex,
pdfauthor={Roberto Velasco-Segura},
pdftitle={Finite Volume Nonlinear Acoustic Propagation - \MDversion},
pdfsubject={Computational Acoustics},
pdfkeywords={Nonlinear Acoustics Simulation GPU Finite Volume},
pdfproducer={Latex with hyperref},
pdfcreator={pdflatex}]{hyperref}

\begin{document}

\title[Finite Volume Nonlinear Acoustic Propagation]{A Finite Volume Approach for the Simulation of Nonlinear Dissipative Acoustic Wave Propagation}

\author{Roberto Velasco-Segura}
\author{Pablo L. Rend\'{o}n}
\email[Author to whom correspondence should be addressed.
Electronic mail: ]{pablo.rendon@ccadet.unam.mx}

\affiliation{Grupo de Ac\'{u}stica y Vibraciones, Centro de Ciencias Aplicadas y Desarrollo Tecnol\'{o}gico, Universidad Nacional Aut\'{o}noma de M\'{e}xico, Ciudad Universitaria--M\'{e}xico, D.F. 04510 M\'{e}xico}

\pacs{43.25.Cb, 43.58.Ta}

\date{12 May 2015}
\thanks{DOI: \url{http://dx.doi.org/10.1016/j.wavemoti.2015.05.006}}

\begin{abstract}
  A form of the conservation equations for fluid dynamics is presented, deduced using slightly less restrictive hypothesis than those necessary to obtain the Westervelt equation.
  This formulation accounts for full wave diffraction, nonlinearity, and thermoviscous dissipative effects.
  A two-dimensional finite volume method using the Roe linearization was implemented to obtain numerically the solution of the proposed equations.
  In order to validate the code, two different tests have been performed:
  one against a special Taylor shock-like analytic solution,
  the other against published results on a High Intensity Focused Ultrasound (HIFU) system,
  both with satisfactory results.
  The code, available under an open source license, is written for parallel execution on a Graphics Processing Unit (GPU), thus improving performance by a factor of over 60 when compared to the standard serial execution finite volume code CLAWPACK 4.6.1, which has been used as reference for the implementation logic as well.
\end{abstract}

\maketitle

\section{Introduction}
\label{sec:introduction}

The Westervelt equation is a classical model for nonlinear acoustic propagation.
It was originally obtained in 1963 by P. J. Westervelt \citep{westervelt} and it describes acoustic propagation taking into account the competing effects of nonlinearity and attenuation.
This and other nonlinear models for acoustics can be obtained adding hypotheses, commonly in the form of restrictions, to the conservation principles of mass, momentum, and energy. 
Two other classical nonlinear acoustics models are the KZK and Burgers equations, and both can be obtained adding restrictions to the Westervelt equation:
to propagation at small angles from a certain axis (quasi-planar propagation) in the case of the KZK equation, 
and to propagation strictly along a single axis (planar propagation) for the Burgers equation \citep{hamilton1998model}.

With a few notable exceptions, solutions for these nonlinear equations, when known, can only be expressed in non trivial forms, and tools like numerical methods are often required to investigate their nature.
Numerical methods, and the means required to implement them, have been in continuous development in recent years.
Primarily, early publications have been devoted to the description of the more restricted models, like the plane Burgers equation \citep{crighton1979asymptotic}, whose exact solution has known analytical expressions \citep{crighton1979asymptotic, blackstock1966connection, jordan}.
Nevertheless, numerical methods in this case have certainly played an important role \citep{yang1992comparative}.
After that, the KZK equation became a widely used model for diagnostic and therapeutic medical applications \citep{duck}, and most of the known solutions have been obtained only by numerical means \citep{khokhlova2001numerical}, including some more recent extensions of the model where the restrictions in propagation direction have been partly relaxed \citep{yuldashev2011simulation, dagrau2011acoustic, varray2011fundamental, varslot2005computer}.
Modern medical applications, such as extracorporeal shock wave therapy \citep{fagnan2008high}, focus control of high intensity focused ultrasound (HIFU) in heterogeneous media \citep{okita2011development}, and ultrasound imaging \citep{huijssen2010iterative}, are now demanding more sophisticated solutions to describe systems where the geometric complexity of the nonlinear acoustic field is important.
Thus, in recent years, a number of schemes have been produced concerned with implementing methods which are not limited in terms of the propagation direction \citep{christopher1991new, albin, demi2011contrast, okita2011development, huijssen2010iterative, karamalis, hallaj1999fdtd, pinton2009heterogeneous, jing2012k, lemoine2013high}, as required in order to solve the Westervelt or Kuznetsov equations, the latter being a model even less restrictive than the Westervelt equation \cite{clason2009boundary}.
These numerical methods are sometimes referred to as {\em full wave methods} \citep{hallaj1999fdtd}.
To the best of our knowledge, no general analytic solutions are known for the full wave case either for the Kuznetsov or the Westervelt equations.
In the present work we aim to give a full wave numerical solution to a set of conservation laws, obtained using slightly less restrictive hypotheses than those necessary to arrive at the Westervelt equation.

A great number of numerical methods have been used to solve the nonlinear acoustic field, some of them operating over the time domain \citep{karamalis, hallaj1999fdtd, okita2011development, pinton2009heterogeneous, lemoine2013high}, while others involve calculations over the frequency domain \citep{christopher1991new, albin, varray, varray2011fundamental, demi2011contrast, jing2012k}.
The numerical method implemented in the present work is a finite volume method, a time domain method.
These methods are based on conservation laws, giving them from the start an intrinsic relation to the equations that conform the basis of all acoustic wave models.
In the present work the CLAWPACK \citep{clawpack} 4.6.1 serial code, which serves as a standard for finite volume schemes, has been used as a reference for the implementation logic in the presented open source C++/CUDA code, which executes the finite volume method in a GPU graphic card, and notably improves the performance compared to serial schemes.
Whereas in the recent literature it is a common practice to use parallelized code for this kind of simulations, this code is mainly run through clusters \citep{albin,  huijssen2010iterative, yuldashev2011simulation, okita2011development, varslot2006forward, pinton2009heterogeneous, terrel2012manyclaw}, and GPU execution is just starting to be used \citep{varray, cudaclaw, karamalis, terrel2012manyclaw}.

The paper is organized as follows:
the relevant equations for this study are described in Section~\ref{sec:equations};
the numerical procedure is described in Section~\ref{sec:numerical};
validation tests for the numerical method, and details of their implementation, are given in Section~\ref{sec:results};
finally, discussion and conclusions are presented in Section~\ref{sec:disc-concl}.

\section{Nonlinear Acoustic Equations}
\label{sec:equations}

In standard form, the Westervelt equation is given as \citep{hamilton1998model}
\begin{align}
  \label{eq:Westervelt}
  \nabla^2 p'
  - \frac{1}{c_0^2}\frac{\partial ^2 p'}{\partial t^2}
  + \frac{\delta}{c_0^4} \frac{\partial ^3 p'}{\partial t^3}
  =
  \frac{-\beta}{\rho_0 c_0^4} \frac{\partial ^2 (p')^2}{\partial t^2}
  \ \text{,}
\end{align}
where $p'$ is the acoustic perturbation pressure, $\nabla^2$ is the Laplacian for spatial variables, $t$ is time, $c_0$ is speed of sound for small signals at an equilibrium state denoted with the zero subscript, $\beta$ is the coefficient of nonlinearity \citep{beyer1998parameter}, and $\delta$ is sound diffusivity \citep{lighthill}.

Since we want to use a finite volume numerical approach, we need to express the Westervelt equation as a system of conservation laws, or more precisely, we need a set of conservation laws as consistent as possible with the Westervelt equation.
To begin with, consider the conservation equations for mass and momentum in a compressible fluid, as stated by Hamilton and Morfey \citep{hamilton1998model},
\begin{align}
  \label{eq:HM:mass}
  \frac{\partial \rho}{\partial t} + \nabla \cdot
  \left(\rho \mathbf{u}\right)& = 0
  \ \text{,} \\
  \label{eq:HM:mom}
  \rho \frac{D  \mathbf{u}}{Dt}
  + \nabla p
  & =
  \mu\nabla^2\mathbf{u}
  +
  \left(\mu_B + \frac{1}{3}\mu\right)\nabla(\nabla\cdot\mathbf{u})
  \ \text{,}
\end{align}
where $p$ is the total pressure, $\rho$ is the total mass density, $\mathbf{u}$ is the fluid velocity (null value of the equilibrium state is assumed), $\mu$ is the dynamic viscosity, and $\mu_B$ is the bulk viscosity.
As we have mentioned before, solving these equations, even numerically, requires some assumptions to be made, in this case because the number of variables is greater than the number of relations among them.
To keep the mentioned consistency, the hypotheses used here are the same as those used by Hamilton and Morfey \citep{hamilton1998model} to obtain the Westervelt equation, with one exception, which we discuss below.
One of these restrictions has to do with the size of the perturbations, $p'$, $\rho'$, and $T'$, considered small and of the same order:
\begin{align*}
  \frac{\rho'}{\rho_0},\frac{p'}{p_0},\frac{T'}{T_0}=O(\epsilon)
  \ \text{,}
\end{align*}
where $\epsilon=|u|/c_0$ is the Mach number, $T$ refers to temperature, and $p_0$, $\rho_0$, and $T_0$, are reference values for pressure, density, and temperature, respectively, so that $p=p_0+p'$, $\rho=\rho_0+\rho'$, and $T=T_0+T'$.
In addition, the fluid is assumed to be irrotational, we use $\rho\frac{D  \mathbf{u}}{Dt} = \frac{\partial \rho\mathbf{u}}{\partial t} + \nabla\cdot(\rho\mathbf{u}\otimes\mathbf{u})$ to stress the conservative character of equation \eqref{eq:HM:mom}, and then we neglect a third order term $\rho'\mathbf{u}\otimes\mathbf{u}$.
In the previous expressions $\otimes$ denotes outer product.
Then, the equation for momentum conservation takes the form
\begin{align}
  \label{eq:ap:mom2}
  \frac{\partial \rho \mathbf{u}}{\partial t} +
  \nabla\cdot\left(\rho_0\mathbf{u}\otimes\mathbf{u}
  + \mathbf{I}p \right)
  & =
  \left(\frac{4}{3}\mu + \mu_B\right)\nabla^2\mathbf{u}
  \ \text{,}
\end{align}
where $\mathbf{I}$ is a $3\times3$ unitary matrix.
Finally, we only need to express $p$ as function of $\rho$ and $\mathbf{u}$, and substitute this expression in \eqref{eq:ap:mom2}. 
Our obtention of this relation follows Hamilton and Morfey's \citep{hamilton1998model} derivation of their equation (40).
Firstly, an expression for the energy conservation is needed, in this case
\begin{align}
  \label{eq:cons:ener}
  \rho_0 T_0 \frac{\partial s'}{\partial t} &= \kappa \nabla^2 T'
  \ \text{,}
\end{align}
written in terms of $s'$, the perturbation for the entropy per unit mass, in such a way that only the linear acoustic mode term remains.
Likewise, two equations of state are proposed, $p=p(\rho,s')$ and $T=T(\rho,s')$, both expressed as Taylor series, which has been chosen to second order for the former and to first order for the latter:
\begin{align}
  \nonumber
  p' &=
  c_0^2 \rho' + c_0^2 \frac{1}{\rho_0} \frac{B}{2A}(\rho')^2 +
  \left(\frac{\partial p}{\partial s'}\right)_{\rho,0} s' \\
  \label{eq:EOS:p:1}
  & \qquad
  + \left(\frac{\partial ^2 p}{\partial \rho\partial s'}\right)_0 s' \rho'
  + \frac{1}{2} \left(\frac{\partial ^2 p}{\partial (s')^2}\right)_{\rho,0} (s')^2
  \ \text{,} \\
  \label{eq:EOS:T:1}
  T' &= \left(\frac{\partial T}{\partial \rho}\right)_{s',0} \rho'
  + \left(\frac{\partial T}{\partial s'}\right)_{\rho,0} s'
  \ \text{,}
\end{align}
where $c_0$, $A$, and $B$ have the usual definitions \citep{beyer}, and subscript $0$ refers to evaluation at equilibrium state,
\begin{align}
  c_0^2 &= \left(\frac{\partial p}{\partial \rho}\right)_{s',0}
  \ \text{,} \\
  A &= \rho_0 \left(\frac{\partial p}{\partial \rho}\right)_{s',0}
  \ \text{,} \quad
  B = \rho_0^2 \left(\frac{\partial ^2 p}{\partial \rho^2}\right)_{s',0}
  \ \text{.}
\end{align}
However, for a weakly thermoviscous fluid, vorticity and thermal modes can be neglected in regions at least on the order of a wavelength away from the solid surfaces, and over such regions the entropy perturbations $s'$ are order $\epsilon^2$, which allows us to neglect the cross term ($\epsilon s'$) in eq. \eqref{eq:EOS:p:1}, and the linear term $s'$ in eq. \eqref{eq:EOS:T:1}:
\begin{align}
  \label{eq:EOS:p:2}
  p' &=
  c_0^2 \rho' + c_0^2 \frac{1}{\rho_0} \frac{B}{2A}(\rho')^2 +
  \left(\frac{\partial p}{\partial s'}\right)_{\rho,0} s'
  \ \text{,} \\
  \label{eq:EOS:T:2}
  T' &= \left(\frac{\partial T}{\partial \rho}\right)_{s',0} \rho'
  \ \text{.}
\end{align}
Also required are the following linear relations: the first order equation of state for $p$,
\begin{align}
  \label{eq:EOSlin}
  p &= c_0^2 \rho
  \ \text{,}
\end{align}
the linearized version of the continuity equation,
\begin{align}
  \label{eq:LMcons}
  \nabla\cdot\mathbf{u} &= \frac{-1}{\rho_0}\frac{\partial \rho'}{\partial t}
  \ \text{,}
\end{align}
and the linear inviscid wave equation for temperature,
\begin{align}
  \label{eq:waveT}
  \nabla^2 T' &= c_0^{-2} \frac{\partial ^2T'}{\partial t^2} \ \text{.}
\end{align}
Finally, we will also use the thermodynamic property
\begin{align}
  \label{eq:termProp}
  c_v - c_p = \frac{TV\alpha^2}{N\kappa}
  \ \text{,}
\end{align}
where $c_v$ and $c_p$ are molar heat capacities at constant volume and pressure respectively, $\kappa$ is isothermal compressibility, and $\alpha$ is the coefficient of thermal expansion, and the following Maxwell relations, taken from  Callen \cite{callen2006thermodynamics},
\begin{align}
  \left(\frac{\partial p}{\partial T}\right)_V &= \left(\frac{\partial s'}{\partial V}\right)_T \ \text{,} 
  \left(\frac{\partial p}{\partial s'}\right)_V = -\left(\frac{\partial T}{\partial V}\right)_{s'} \ \text{,} 
  \nonumber \\ 
  \left(\frac{\partial V}{\partial T}\right)_p &= -\left(\frac{\partial s'}{\partial p}\right)_T \ \text{.}
  \label{eq:MaxRel}
\end{align}
Substituting expressions \eqref{eq:cons:ener}, \eqref{eq:EOS:T:2}, \eqref{eq:EOSlin}, \eqref{eq:LMcons}, \eqref{eq:waveT}, \eqref{eq:termProp}, and \eqref{eq:MaxRel} into \eqref{eq:EOS:p:2}, we get, after some algebra, the desired relation
\begin{align}
  \label{eq:finalp}
  p = c_0^2 \rho + c_0^2 \frac{1}{\rho_0} \frac{B}{2A}(\rho')^2 +
  \frac{\kappa}{\rho_0}
  \left(
    \frac{1}{c_v} - \frac{1}{c_p}
  \right)
  \rho_0 \nabla\cdot\mathbf{u}
  \ \text{,}
\end{align}
which is indeed equivalent to eq. (40) in Hamilton and Morfey \citep{hamilton1998model}.

Now, in order to get the desired system of conservation laws, we substitute eq. \eqref{eq:finalp}, which combines the equations of state and energy conservation, into \eqref{eq:ap:mom2}, leading to the appearance of three terms: a linear term, a nonlinear term, and a dissipative term that happens to complete the coefficient of sound diffusivity as presented by Lighthill \citep{lighthill}:
\begin{align}
\label{eq:def-delta}
  \delta = \frac{1}{\rho_0}
  \left(
    \frac{4}{3}
    \mu+\mu_B
  \right)
  +
  \frac{\kappa}{\rho_0}
  \left(
    \frac{1}{c_v}
    -\frac{1}{c_p}
  \right)
  \ \text{.}
\end{align}

Thus, the resulting equations, after substitution of \eqref{eq:finalp} in \eqref{eq:ap:mom2}, and which are the base for our finite volume implementation, are
\begin{subequations}
\label{eq:OurSystem}
\begin{align}
  \label{eq:final:cons:mass}
  \frac{\partial \rho}{\partial t} + \nabla \cdot
  \left(\rho \mathbf{u}\right)& = 0
  \ \text{,} \\
  \label{eq:final:cons:mom}
  \frac{\partial \rho \mathbf{u}}{\partial t} +
  \nabla\cdot\Bigg[
    \rho_0 \mathbf{u}\otimes\mathbf{u} 
    \qquad \qquad \qquad \qquad
  \nonumber \\
  \qquad \qquad
    + \ \mathbf{I}
    \left(
      c_0^2 \rho + \frac{c_0^2}{\rho_0} (\beta-1)(\rho-\rho_0)^2
    \right)
  \Bigg]
  &=
  \rho_0 \delta \nabla^2 \mathbf{u}
  \ \text{,} 
\end{align}
\end{subequations}
where $\beta = 1 + \frac{B}{2A}$ is the nonlinear coefficient \citep{beyer1998parameter}.
Since the only physical variables to appear in equations \eqref{eq:OurSystem} are $\rho$ and $\mathbf{u}$, it would seem natural to use the equations of state, \eqref{eq:EOS:p:2} and \eqref{eq:EOS:T:2}, to recover the pressure, but they cannot be used directly for this purpose because the variables $s'$ and $T$ appear explicitly in the equations of state and they are absent from equations \eqref{eq:OurSystem}.
A similar situation is encountered when trying to use equation \eqref{eq:finalp} for this same purpose: the values of $c_v$, $c_p$  and $\kappa$ are not necessarily known. We opt here for the simplest solution to this problem, which is to use the linear approximation $p=c_0^2 \rho$ to obtain the pressure $p$.
This is generally acknowledged to be a good approximation in this context because nonlinear profile distortion, which will clearly affect $p$ as well, is associated with the presence of nonlinear terms in equations \eqref{eq:OurSystem}, and is caused by cumulative, not local effects.
If we wished to take local nonlinear effects into account as well, then we would need to modify this last expression so as to include higher-order terms in $\rho$ as well.

It is important to note that during the obtention of equations \eqref{eq:OurSystem} it was not necessary to drop the second-order Lagrangian density term
$\rho_0 u^2 / 2 - p^2 /( 2 \rho_0 c_0^2)$ as Hamilton and Morfey \citep{hamilton1998model} have done in their derivation of the Westervelt equation.
As a result, we may consider equations \eqref{eq:OurSystem} to constitute a model slightly more general than the Westervelt equation, which could actually be used to evaluate the importance of that term in appropriate conditions.
This investigation is left for future work. 
Note that equations like \eqref{eq:final:cons:mom} are sometimes called {\em balance laws} because of the presence of the source term \citep{leveque-book-fv}, identified above as the right hand side of the equation.

\section{Numerical Method}
\label{sec:numerical}

For the rest of this paper only two spatial dimensions are considered.
The theory presented in Section~\ref{sec:equations} is valid for three spatial dimensions, but we have chosen to restrict our implementation, for the moment, to two dimensions, mainly to keep down computational costs.

Using a variable renormalization
\begin{align}
  \label{eq:renorm}
  \frac{t c_0}{L}  \mapsto t
  \ \text{,} \qquad
   \frac{x_j}{L} \mapsto  x_j
  \ \text{,} \quad \text{with} \quad
  j = 1,2
\end{align}
where $L$ is a typical length in the system, the equations \eqref{eq:OurSystem} can be rewritten in non-dimensional differential conservation law form \citep{sharma2010quasilinear}, as follows:
\begin{align}
  \label{eq:OurSysAdim}
  \frac{\partial q}{\partial t} + \frac{\partial f_1}{\partial x_1} + \frac{\partial f_2}{\partial x_2} = \psi \ \text{,}
\end{align}
where $q$ is the vector of conserved quantities, $f_j$ are vectors containing their fluxes (in $x_j$ direction), and $\psi$ is a source term.
These vectors are written in the following form:
\begin{align}
  q
  &=
  \left(
    \begin{array}{c}
      q^1 \\
      q^2 \\
      q^3
    \end{array}
  \right)
  =
  \frac{\rho}{\rho_0}
  \left(
    \begin{array}{c}
      1 \\
      u^1/c_0 \\
      u^2/c_0
    \end{array}
  \right) \ \text{,}
  \label{eq:adim:q}
  \\
  f_j
  &= \frac{q^{j+1}}{(q^1)^2}
  \left(
    \begin{array}{c}
      (q^1)^2 \\
      q^2 \\
      q^3
    \end{array}
  \right)
  + \phi
  \left(
    \begin{array}{c}
      0 \\
      \delta_{1j} \\
      \delta_{2j}
    \end{array}
  \right) \ \text{,}
  \nonumber
  \\
  \phi &= q^1 + (\beta-1)(q^1-1)^2 \ \text{,}
  \nonumber
  \\
  \psi
  &=
  \tilde{\delta}
  \left(
    \begin{array}{c}
      0 \\
      \nabla^2 q^2/q^1 \\
      \nabla^2 q^3/q^1
    \end{array}
  \right) \ \text{,}
  \nonumber
\end{align}
where $\delta_{ij}$ is a Kronecker delta. Following LeVeque \citep{leveque-book-fv}, superscripts have been used to denote vector components; when powers are needed for those variables, an extra set of parentheses is added, {\em e.g.} $(q^1)^2$.
The coefficient
\begin{align}
  \label{eq:phi-tdelta}
  \tilde{\delta} = \frac{\delta}{c_0 L}
\end{align}
is a dimensionless form of the sound diffusivity $\delta$ as given by Lighthill \citep{lighthill},
and the parameter $\beta$ is the coefficient of nonlinearity \citep{beyer1998parameter}.

In order to implement a fractional step method, we use the following decomposition
\begin{align}
  \label{eq:fs:cl1}
  \frac{\partial q}{\partial t} + \frac{d f_1}{d q}\frac{\partial q}{\partial x_1} &= 0 \ \text{,}\\
  \label{eq:fs:cl2}
  \frac{\partial q}{\partial t} + \frac{d f_2}{d q}\frac{\partial q}{\partial x_2} &= 0 \ \text{,}\\
  \label{eq:fs:source}
  \frac{\partial q}{\partial t} &= \psi \ \text{,}
\end{align}
where $\frac{d f_j}{d q}$ is a $3\times 3$ Jacobian matrix.
The fractional step method in this case consists in applying consecutive numerical schemes for equations \eqref{eq:fs:cl1}, \eqref{eq:fs:cl2}, and \eqref{eq:fs:source}.
Since \eqref{eq:fs:cl1} and \eqref{eq:fs:cl2} are essentially the same equation, switching positions $2$ and $3$ in every vector, they can be solved numerically with the same scheme.
For this reason, we have dropped the $j$ subscript in the following sections, and we work with equation \eqref{eq:fs:cl1} only.
To obtain a numerical solution for expression \eqref{eq:fs:source}, a standard central finite difference second-order method was used.

\subsection{Finite Volume Method}
\label{sec:finite-volume-method}

A numerical solution to equation \eqref{eq:fs:cl1} is sought using a finite volume high-resolution method \citep{leveque-book-fv} of the form
\begin{align}
  \nonumber
  Q_i^{n+1} &= Q_i^n
  -\frac{\Delta t}{\Delta x}
  \left(
    \mathcal{A}^-\Delta Q_{i+1/2}
    + \mathcal{A}^+\Delta Q_{i-1/2}
  \right) \\
  \label{eq:fv-hr}
  & \quad
  -\frac{\Delta t}{\Delta x}
  \left(
    \tilde{F}_{i+1/2}
    -\tilde{F}_{i-1/2}
  \right)
  \ \text{,}
\end{align}
where, as usual in finite volume methods,
\begin{align*}
  Q^n_i \approx \frac{1}{\Delta x} \int_{x_{i-1/2}}^{x_{i+1/2}}
  q(x,t_n) \ dx
  \ \text{.}
\end{align*}
Expression \eqref{eq:fv-hr}, is in principle, a numerical method formulated for a linear system of conservation laws.
The procedure needed to use it with a nonlinear system is briefly described below, and further details can be found in LeVeque's textbook \citep{leveque-book-fv}.
Expression \eqref{eq:fv-hr} approximates the new values of $q$ in the cell $i$, that is $Q^{n+1}_i$, as a function of the old value $Q^n_i$ and the fluxes from neighboring cells.
These fluxes are described by the matrix $\frac{d f(q)}{d q}$, which is dependent on $q$ because of the nonlinear character of equation \eqref{eq:OurSysAdim}, and is thus dependent on position.
This matrix is evaluated at $i-1/2$ to determine the flux between cell $i$ and cell $i-1$, and at $i+1/2$ to determine the flux between cell $i+1$ and cell $i$.
We have implemented a Roe linearization \citep{roe, leveque-book-fv} scheme to approximate this matrix $\frac{d f(q)}{d q}$ at these positions.
Once we have these approximations, say for $i-1/2$, their eigenvalues $s_{i-1/2}^p$ and eigenvectors $r_{i-1/2}^p$ can be obtained.
Further details are presented in Section~\ref{sec:roe}.

The second term in expression \eqref{eq:fv-hr} is constructed with terms sometimes called {\em fluctuations} \citep{leveque-book-fv},
\begin{align}
  \label{eq:fluc:m}
  \mathcal{A}^-\Delta Q_{i+1/2}
  &= \sum_{p=1}^3(s_{i+1/2}^p)^- \mathcal{W}_{i+1/2}^p
  \ \text{,}\\
  \label{eq:fluc:p}
  \mathcal{A}^+\Delta Q_{i-1/2}
  &= \sum_{p=1}^3(s_{i-1/2}^p)^+ \mathcal{W}_{i-1/2}^p
  \ \text{,}
\end{align}
where
\begin{align*}
  \mathcal{W}^p_{i-1/2} = \alpha^p_{i-1/2} r^p_{i-1/2}
\end{align*}
and $\alpha^p_{i-1/2}$ just represent a normalization of the eigenvectors $r^p_{i-1/2}$ in order to satisfy
\begin{align*}
  Q_i - Q_{i-1} = \sum_{p=1}^3 \alpha_{i-1/2}^p r^p_{i-1/2}
  \ \text{.}
\end{align*}
Observe that superscripts are just an index for the eigenvector decomposition, and they do not represent any algebraic operation.

The definitions in equations \eqref{eq:fluc:m} and \eqref{eq:fluc:p} correspond to a Godunov first-order method.
What this means is that if the third term of equation \eqref{eq:fv-hr} were absent -- and as it is shown below this could eventually happen -- then equation \eqref{eq:fv-hr}, because of its second RHS term, would have the form of a Godunov method.

The third term in the RHS of expression \eqref{eq:fv-hr} is a high resolution correction term \citep{leveque-book-fv}, of the form
\begin{align}
  \label{eq:limiter}
  \tilde{F}_{i-1/2}
  =
  \frac{1}{2}
  \sum_{p=1}^3
  \left|s_{i-1/2}^p\right|
  \left(
    1-
    \frac{\Delta t}{\Delta x}
    \left|s_{i-1/2}^p\right|
  \right)
  \stackrel{\sim}{\smash{\mathcal{W}}\rule{0pt}{1.3ex}}^p_{i-1/2}
  \ \text{,}
\end{align}
where
\begin{align*}
  \stackrel{\sim}{\smash{\mathcal{W}}\rule{0pt}{1.3ex}}^p_{i-1/2}
  &= \tilde{\alpha}^p_{i-1/2}r^p_{i-1/2}
  \ \text{,}
  \\
  \tilde{\alpha}_{i-1/2}^p
  &= \alpha_{i-1/2}^p \phi( \theta_{i-1/2}^p )
  \ \text{,}
  \\
  \phi(\theta_{i-1/2}^p) &=
  \max(0, \min((1 + \theta_{i-1/2}^p )/2, 2, 2\theta_{i-1/2}^p))
  \ \text{,}
  \\
  \theta_{i-1/2}^p
  &=
  \frac{\mathcal{W}_{I-1/2}^p\cdot\mathcal{W}_{i-1/2}^p}
  {\mathcal{W}_{i-1/2}^p\cdot\mathcal{W}_{i-1/2}^p}
  \ \text{,}
  \\
  I &=
  \begin{cases}
    i-1 & \text{if } s_{i-1/2}^p > 0  \ \text{,}\\
    i+1 & \text{if } s_{i-1/2}^p < 0  \ \text{.}
  \end{cases}
\end{align*}
When substituting expressions \eqref{eq:fluc:m}, \eqref{eq:fluc:p}, and \eqref{eq:limiter} into equation \eqref{eq:fv-hr}, it is noticeable that equation \eqref{eq:fv-hr} is a sum over families of waves indexed with $p$.
This decomposition is just an eigenvector representation of the system.

In the case $\phi=1$, equation \eqref{eq:limiter} corresponds to a Lax-Wendroff method, which is a second order method but presents spurious oscillations near edges.
Function $\phi$ as described above is a standard flux limiter, in this case known as {\em monotonized central-difference} \citep{leveque-book-fv}.
This limiter can take values between $0$ and $2$, and approaches $1$ when the solution corresponding to the $p$-th family is smooth.
Since in this case the limiter $\phi$ is TVD, it avoids spurious oscillations and retains, to a good extent, the accuracy of the second order method in smooth regions.
The above mentioned smoothness is evaluated through the coefficient $\theta_{i-1/2}^p$, which is a normalized projection of $\mathcal{W}_{I-1/2}^p$ over $\mathcal{W}_{i-1/2}^p$.
If both are close to each other, then $\theta_{i-1/2}^p \approx 1$, as they will have approximately the same size and direction.

\subsection{Roe Linearization}
\label{sec:roe}

As we have previously said, we need a matrix $\hat{A}_{i-1/2}$ which approximates $\frac{d f}{d q}$ at the middle point between $i$ and $i-1$, for any given $i$.
Since $f$ is a nonlinear function of $q$, it is expected that $\hat{A}_{i-1/2}$ will be a function of $Q_i$ and $Q_{i-1}$.
For the purpose of linearization this matrix is considered constant in a finite volume method constructed for the linear case, as in expression \eqref{eq:fv-hr}.

Two properties are expected of such approximations \citep{leveque-book-fv}:
\begin{description}
\item[(i)] consistency, {\em i.e.} convergence to $\frac{d f}{d q}$ when $Q_i~\rightarrow~Q_{i-1}$; and
\item[(ii)] diagonalizability by means of real eigenvalues.
\end{description}
The Roe linearization is a technique designed to find a matrix $\hat{A}_{i-1/2}$ which, besides the previous properties, satisfies
\begin{align}
  \label{eq:RoeCond}
  f(Q_i) -f(Q_{i-1}) = \hat{A}_{i-1/2} (Q_i  - Q_{i-1})
  \ \text{,}
\end{align}
which is a useful property because it ensures the method will be conservative.
Furthermore, for the shock wave cases, where there is a region with a strong change for a single $p$ wave family, {\em i.e.} much bigger than the changes in other wave families, expression \eqref{eq:RoeCond} implies this $p$-wave is (in the limit) an eigenvector of $\hat{A}_{i-1/2}$, which is the case except for the small regions when two shocks collide \citep{leveque-book-fv}.

Now, using a variable change $z=z(q)$, which fortuitously coincides with dimensionless primitive variables,
\begin{align*}
  \left(
  \begin{array}{c}
    z^1 \\
    z^2 \\
    z^3
  \end{array}
  \right)
  =
  \left(
    \begin{array}{c}
      q^1 \\
      q^2 / q^1 \\
      q^3 / q^1
    \end{array}
  \right)
  =
  \left(
    \begin{array}{c}
      \rho / \rho_0 \\
      u^1 / c_0 \\
      u^2 / c_0
    \end{array}
  \right)
  \ \text{,}
\end{align*}
to define a path
\begin{align*}
  \hat{z}(\xi) = Z_{i-1} + (Z_i - Z_{i-1}) \xi \ \text{,}
  \qquad
  \text{where}
  \qquad
  Z_i=z(Q_i) \ \text{,}
\end{align*}
and performing the integrals \citep{leveque-book-fv},
\begin{align*}
  \int_0^1 \frac{f(q(\hat{z}))}{d\hat{z}} \ d\xi
  \ \text{,} \quad
  \int_0^1 \frac{q(\hat{z})}{d\hat{z}} \ d\xi
  \ \text{,}
\end{align*}
the desired matrix, satisfying all of the properties mentioned above, is found: %
{\renewcommand{\arraystretch}{2.5}
\begin{align*}
  \hat{A}_{i-1/2}
  =
  \left(
    \begin{array}{ccc}
      0 & 1 & 0 \\
      -2\dfrac{(\bar{Z}^{2})^2}{\bar{Z}^{1}} + \Phi
      & 2\dfrac{\bar{Z}^{2}}{\bar{Z}^{1}} & 0  \\
      -2\dfrac{\bar{Z}^{2}\bar{Z}^{3}}{\bar{Z}^{1}} & \dfrac{\bar{Z}^{3}}{\bar{Z}^{1}}
      & \dfrac{\bar{Z}^{2}}{\bar{Z}^{1}}
    \end{array}
  \right)
\end{align*}}
where
\begin{align*}
  \bar{Z}^{p} = \frac{1}{2}\left(Z^p_{i} + Z^p_{i-1}\right)
  \quad \text{and} \quad
  \Phi = 1 + 2(\beta-1)(\bar{Z}^{1}-1) \ \text{.}
\end{align*}
Again, superscripts for $z$, $Z$, $\bar{Z}^{}$, $s$, $q$ and $u$, are indices, powers of these variables are denoted with an extra pair of parentheses.
The eigenvalues of this matrix are
\begin{align}
  \label{eq:lambda1}
  s_{i-1/2}^1 &= \frac{\bar{Z}^{2}}{\bar{Z}^{1}}
  + \sqrt{\frac{(\bar{Z}^{2})^2}{(\bar{Z}^{1})^2}
    - 2 \frac{(\bar{Z}^{2})^2}{\bar{Z}^{1}}
    + \Phi
  } \ \text{,}\\
  \label{eq:lambda2}
  s_{i-1/2}^2 &= \frac{\bar{Z}^{2}}{\bar{Z}^{1}}
  - \sqrt{\frac{(\bar{Z}^{2})^2}{(\bar{Z}^{1})^2}
    - 2 \frac{(\bar{Z}^{2})^2}{\bar{Z}^{1}}
    + \Phi
  } \ \text{,}\\
  \label{eq:lambda3}
  s_{i-1/2}^3 &= \frac{\bar{Z}^{2}}{\bar{Z}^{1}}
\end{align}
and the respective eigenvectors can be obtained in straightforward fashion.
When
\begin{align}
  \label{eq:Hyp:Prim}
  Q_i=Q_{i-1}\approx q
\end{align}
is used, and these eigenvalues are expanded as Taylor series to first order, more familiar expressions are obtained,
\begin{align}
  \label{eq:lambda:p1}
  s_{i-1/2}^1 &= \frac{u^1+c}{c_0}
  +O(\epsilon^2)
  \ \text{,}\\
  \label{eq:lambda:p2}
  s_{i-1/2}^2 &= \frac{u^1-c}{c_0}
  +O(\epsilon^2)
  \ \text{,}\\
  \label{eq:lambda:p3}
  s_{i-1/2}^3 &= \frac{u^1}{c_0}
  +O(\epsilon^2)
  \ \text{,}
\end{align}
where the speed of sound $c$ has been calculated from equation \eqref{eq:finalp} as
\begin{align*}
  c = \sqrt{\frac{\partial p}{\partial \rho}}
  =c_0\left(1+\frac{\rho'}{\rho_0}(\beta-1)\right)+O(\epsilon^2)
  \ \text{.}
\end{align*}
Considering the non-dimensional character of the variables used here, expressions \eqref{eq:lambda:p1}, \eqref{eq:lambda:p2} and \eqref{eq:lambda:p3}, are consistent with other finite volume methods for Euler equations \citep{leveque-book-fv}.

\begin{figure}[htb]
\centering
\framebox{
  \resizebox{3.3in}{!}{
    \includegraphics{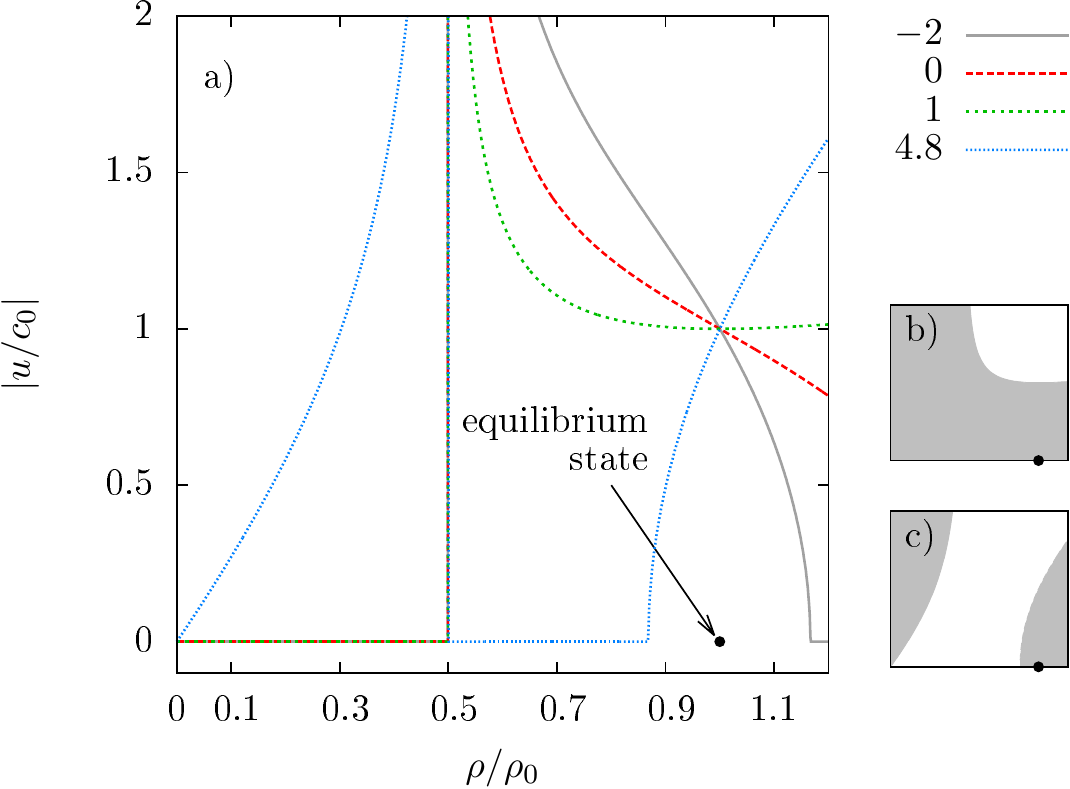}
  }
}
\caption{a) Contours $|u/c_0|=\sqrt{S(\rho/\rho_0)}$ which delimit regions where eigenvalues given by \eqref{eq:lambda1} and \eqref{eq:lambda2} are real.
Numbers in labels correspond to values of $\beta$.
Insets b) and c) are examples of shaded regions where eigenvalues are real, for $\beta=1$ and $\beta=4.8$, respectively%
.}
\label{fig:EV}
\end{figure}

A limitation of this method is evidenced by analysis of expressions \eqref{eq:lambda1} and \eqref{eq:lambda2}: eigenvalues can become non-real.
However, the argument of the square roots which appear in these equations will remain positive so long as $\rho/\rho_0 < 0.5$ and $(u/c_0)^2 > S(\rho/\rho_0)$, or $\rho/\rho_0 > 0.5$ and $(u/c_0)^2 < S(\rho/\rho_0)$, where \eqref{eq:Hyp:Prim} has been used, and
\begin{align*}
  S(\rho/\rho_0) =  \max\left\{0,
    \frac{1+2(\beta-1)((\rho/\rho_0)-1)}{2(\rho/\rho_0)-1}
    (\rho/\rho_0)^2
  \right\}
  \ \text{.}
\end{align*}
This is illustrated for some representative values of $\beta$ in figure \ref{fig:EV}.
The permitted regions are above the lines for $\rho/\rho_0<0.5$, and below them for $\rho/\rho_0>0.5$.
The relevant region for our cases is around the equilibrium state $\rho/\rho_0=1$, $u/c_0=0$, where the eigenvalues will remain real provided that perturbations stay small.
In this way, perturbations as large as $10\%$ can be used for typical values $\beta\approx 5$ (see for example Beyer \citep{beyer1998parameter}).
Note that some of the permitted regions for $\rho/\rho_0<0.5$, {\em e.g.} for $\beta=4.8$, are not reachable via a continuous path from the equilibrium state.

This limited region for the allowed values of variables is consistent with the perturbative hypotheses used to obtain equations \eqref{eq:OurSystem}.
Thus, it reflects the limits of the model itself, and not of the numerical method as such.
Indeed, the eigenvalues of $\frac{d f}{d q}$ present the same problem.

As we have pointed out, this numerical method corresponds to equation \eqref{eq:fs:cl1}, but it can be used for \eqref{eq:fs:cl2} as well if the second and third entries of its vectors are switched, so that $\bar{Z}^{3}$ appears in their corresponding eigenvalues instead of $\bar{Z}^{2}$.

\section{Implementation and Validation of the Numerical Method}
\label{sec:results}

The mesh used in all the simulations was Cartesian, with $\Delta x_1 = \Delta x_2 = \Delta x$;
in what follows, only $\Delta x$ is used.
The value of $\Delta t$ was controlled by the CFL value
\begin{align*}
  \nu = \frac{\Delta t}{\Delta x} \max\{s_{i-1/2}^p\}
\end{align*}
where the maximum is taken over the entire mesh, in both directions, and over the three possible values of the index $p$.
For every time step the value of $\Delta t$ was adjusted, using the $\nu$ calculated in the previous step, to maintain $\nu$ around some {\em wished} value $\nu_W$, and smaller than 1; see LeVeque \citep{leveque-book-fv}.

Since all the simulations presented in this work use only wave packets, then it is possible to save computational resources by using a mobile domain.
Following Albin {\em et al.} \citep{albin}, two stages of the simulation were implemented:
in the first stage, from $t=t_0<0$ to $t=t_1>0$, the domain stays still and the boundary condition introduces the propagating wave until it reaches the center of the domain at $t=t_1$;
in the second stage the domain moves with the propagation wave.
Additionally, the time scale was adjusted so that $t=0$ when the center of the packet was precisely on the left border, and the value of $t_0$ was chosen for the packet to be numerically zero at  machine precision.
Making use of this technique, simulations were performed in domains 100 wavelengths long.

To validate the code, two sets of tests were performed.
Firstly, in Section~\ref{sec:agains-analyt-solut}, against a Taylor shock solution \citep{jordan}, which is an analytic plane wave solution.
This  constitutes a test for nonlinearity and diffusivity, but not for full wave diffraction, because of the plane nature of the wave.
As we have previously said, no analytic solutions showing full wave diffraction are known, to our knowledge.
Thus, the second set of tests were performed against another numerical solution, for a HIFU (high intensity focused ultrasound) system, obtained by Albin {\em et al.} \citep{albin}, by a different method, in this case a pseudospectral Fourier continuation method.
The results of this comparison are presented in Section~\ref{sec:agains-anoth-numer}.

The present code is written in C++/CUDA v5.0, and it was executed on a Tesla C2075 GPU graphic card (448 cores and 6GB on RAM), in all cases executing blocks of $16\times16$ threads, installed on a standard PC i3-550 with 16GB on RAM, running Debian/jessie.
The finite volume code used for performance comparison, described in Section~\ref{sec:performance}, was executed on the same computer.
All the simulations presented here were performed with double precision.

\subsection{Validation Against a Plane Wave Analytic Solution}
\label{sec:agains-analyt-solut}

\begin{table*}
    \begin{tabular}{ccccccccc}
      \toprule
      & & \multicolumn{3}{c}{Convergence rate} &
      \multirow{2}{*}{\parbox{1.8cm}{\centering Worst $R^2$ value}} 
      & \multicolumn{2}{c}{Error on $\eta=82$} \\
      \cmidrule(r){3-5} \cmidrule(r){7-8}
      & Error & Best & Worst & Mean & & Best & Worst \\
\midrule
\multirow{2}{*}{$\theta_T=0$} &
$E_1$ &
2.0027 & 1.6288 & 1.7637 & 0.9821 & 2.36e-05 & 2.91e-04 \\
& $E_\infty$ &
1.9043 & 1.4900 & 1.6252 & 0.9727 & 2.96e-04 & 3.58e-03 \\
\midrule
\multirow{2}{*}{$\theta_T=\pi/32$} &
$E_1$ &
1.7435 & 1.1967 & 1.5366 & 0.9892 & 2.70e-04 & 3.78e-04 \\
& $E_\infty$ &
1.4553 & 1.2140 & 1.3652 & 0.9865 & 3.56e-03 & 5.98e-03 \\
\midrule
\multirow{2}{*}{$\theta_T=\pi/16$} &
$E_1$ &
1.5695 & 1.1518 & 1.3994 & 0.9860 & 4.62e-04 & 7.00e-04 \\
& $E_\infty$ &
1.3170 & 1.1840 & 1.2508 & 0.9868 & 6.73e-03 & 8.71e-03 \\
\midrule
\multirow{2}{*}{$\theta_T=\pi/8$} &
$E_1$ &
1.3772 & 1.2776 & 1.3145 & 0.9848 & 7.97e-04 & 1.24e-03 \\
& $E_\infty$ &
1.2136 & 1.1419 & 1.1862 & 0.9820 & 1.06e-02 & 1.30e-02 \\
\midrule
\multirow{2}{*}{$\theta_T=\pi/4$} &
$E_1$ &
1.3652 & 1.2220 & 1.2855 & 0.9902 & 1.31e-03 & 2.14e-03 \\
& $E_\infty$ &
1.2925 & 1.2478 & 1.2682 & 0.9940 & 1.12e-02 & 1.78e-02 \\
      \bottomrule
    \end{tabular}
  \caption{Taylor shock, convergence analysis.}
  \label{tab:error-ts}
\end{table*}

It is straightforward to verify that
\begin{align*}
  \frac{p'}{p_0} &= \frac{-\tilde{\delta}}{\beta} \tanh(x-t) \ \text{,}
\end{align*}
the sometimes called Taylor shock \citep{jordan}, is a traveling wave solution of the dimensionless version of the Westervelt equation \eqref{eq:Westervelt},
\begin{align*}
  \nabla^2
  \left(\frac{p'}{p_0}\right)
  - \frac{\partial ^2 }{\partial t^2}
  \left(\frac{p'}{p_0}\right)
  + \tilde{\delta} \frac{\partial ^3 }{\partial t^3}
  \left(\frac{p'}{p_0}\right)
  =
  -\beta \frac{\partial ^2 }{\partial t^2}
  \left(\frac{p'}{p_0}\right)^2
  \ \text{,}
\end{align*}
where \eqref{eq:renorm} and \eqref{eq:phi-tdelta} were used.
It is then reasonable to expect that a similar expression could be a solution to equations \eqref{eq:OurSysAdim}.
To find this expression, first consider a solution which travels along the $x_1$ direction without deformation $q^1 = q^1(x-t)$, $q^2 = q^2(x-t)$, $q^3 = 0$.
Then equations \eqref{eq:OurSysAdim} reduce to $N[q^1] = 0$, where the operator $N$ is defined as
\begin{align*}
  N[q^1] &= \frac{\partial q^1}{\partial t}
  - \tilde{\delta}
  \frac{\partial ^2}{\partial x_1^2}\left(\frac{q^1-1}{q^1}\right)
  \nonumber \\ & \quad
  + \frac{\partial }{\partial x_1}
  \left(
    \left(\frac{q^1-1}{q^1}\right)^2
    + q^1
    + (\beta-1)(q^1-1)^2
  \right)
  \ \text{.}
\end{align*}
On the other hand, an approximate solution $\tilde{q}^1$ does not satisfy the equation, but yields $N[\tilde{q}^1]\simeq 0$.
After some algebra it can be shown that
\begin{align}
  \label{eq:TSapprox}
  \tilde{q}^1 &= 1 - \frac{\tilde{\delta}}{\beta} \tanh(x-t)
\end{align}
is such that
\begin{align*}
  N[\tilde{q}^1] = O(\epsilon^3)
  \ \text{.}
\end{align*}
Note that, in this case $\tilde{\delta}/{\beta} = O(\epsilon)$, accordingly with the definition stated at the beginning of Section \ref{sec:equations}.
Now, in order to know how far this approximation $\tilde{q}^1$ is from the exact solution $q^1$, consider another approximate solution, adding a constant $\gamma$ to a real solution $q^1$, such that $\gamma = O(\epsilon^n)$,
$n>1$, and $1 \simeq q^1 = O(\epsilon^0)$.
In this case
\begin{align*}
  \tilde{q}^1 = q^1 + \gamma
  \quad 
  \text{implies that} 
  \quad
  N[\tilde{q}^1] = O(\epsilon^{n+1})
  \ \text{.}
\end{align*}
Then, assuming the approximate solution is as smooth as the exact solution, that is, it can be considered locally as the exact solution plus a constant, for the case corresponding to equation \eqref{eq:TSapprox} this constant is $O(\epsilon^2)$, or
\begin{align}
  q^1 = 1 - \frac{\tilde{\delta}}{\beta} \tanh(x-t) + O(\epsilon^2)
  \ \text{.}
  \label{TS:approx:E}
\end{align}
In this section we have used the analytic expression \eqref{TS:approx:E} to measure the difference between the numerical results and the exact solution, which is valid since this difference is at least one order of magnitude larger than $O(\epsilon^2)\simeq 10^{-14}$, as shown below.

For the simulations in this section, all the boundary conditions (left, right, top, and bottom) were taken to be of the form
\begin{align}
  \label{eq:OurTaylorS}
  q^1 = 1 - \frac{\tilde{\delta}}{\beta} \tanh(x-t) \ \text{,} \
  q^2 = q^1 -1 \ \text{,} \
  q^3 = 0
  \ \text{,}
\end{align}
for both stages. As can be seen from equation \eqref{eq:OurTaylorS}, in this case the independent values of $\beta$ and $\tilde{\delta}$ are not important, only their ratio is, and it determines $O(\epsilon)$.
For all following simulations, we have chosen $\tilde{\delta}/\beta = 10^{-7}$, so that $O(\epsilon^2)\simeq 10^{-14}$.

Since at cell scale the simulation fluxes are horizontal or vertical, but not diagonal, the simulation results are expected to depend on the direction of propagation \citep{huijssen2010iterative,pinton2009heterogeneous,lemoine2013high}.
In order to appreciate whether this poses a problem for our code or not, we tested propagation at several propagation angles $\theta_T$, measured from the $x_1$-axis:
$0$, $\pi/32$, $\pi/16$, $\pi/8$, $\pi/4$.
These angles are thought to be representative of most of the relevant possibilities.
Equations \eqref{eq:OurTaylorS}, as they are, correspond to $\theta_T=0$; to use them with a different value of $\theta_T$, we have applied a standard rotation change of variables.

Other important parameters with respect to quality of results are $\nu_W$ and $(\Delta x)^{-1}$.
With regard to the former $\nu_W$, as introduced at the beginning of Section \ref{sec:results}, we recall that its value only approximates the real CFL value $\nu$.
In order to calculate the difference between the two, its standard deviation was measured, with the result that it never exceeded the value $10^{-8}$.
Regarding the latter, for which we use letter $\eta = (\Delta x)^{-1}$, the value of $L$, in equation \eqref{eq:renorm}, was chosen as the distance between points of the wave such that the amplitude at these points is $10\%$ and $90\%$ of the maximum amplitude.
In dimensionless variables this distance then equals $1$, and in this case $\eta$ then represents the number of cells across the shock wave.

\begin{figure}[htb]
\centering
\framebox{
  \resizebox{3.3in}{!}{
    \includegraphics{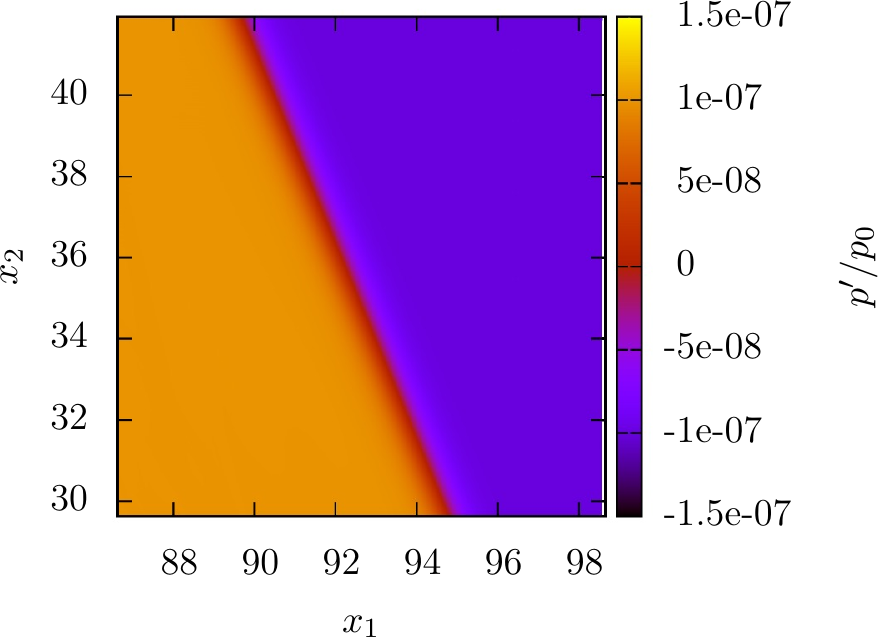}
  }
}
\caption{Taylor shock, view from above, numerical solution for $\theta_T = \pi/8$, $\eta=82$ and $t=100$%
.}
\label{fig:taylor-upper}
\end{figure}
\begin{figure}[htb]
\centering
\framebox{
  \resizebox{3.3in}{!}{
    \includegraphics{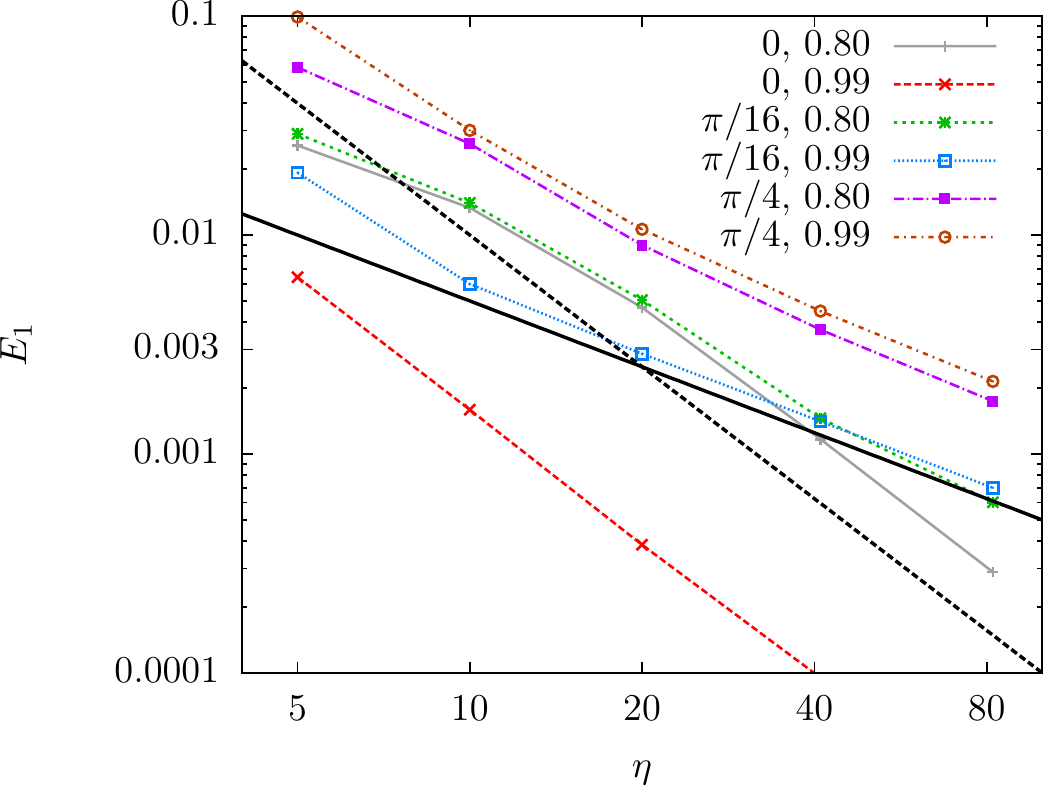}
  }
}
\caption{
Taylor shock, error $E_1$ as a function of $\eta$.
Numbers in labels correspond to values of $\theta_T$, $\nu_W$.
For reference, rates of convergence as $\eta^{-1}$ and $\eta^{-2}$ are indicated by means of thick solid and thick dashed lines, respectively%
.}
\label{fig:taylor-error}
\end{figure}

The simulations reported here were performed for the angles mentioned above and for combinations of the following values: $\eta = 5$, $10$, $20$, $41$, $82$; $\nu_W=0.6$, $0.7$, $0.8$, $0.9$, $0.99$.
Making use of equations \eqref{eq:adim:q} and \eqref{eq:EOSlin} to recover the physical variables, figure~\ref{fig:taylor-upper} shows how a typical simulation looks like, in this case for $\theta_T = \pi/8$, $\eta=82$, and $t=100$, that is, after the wave has traveled a distance $100$ times larger than its width.

Error analysis is depicted in table \ref{tab:error-ts} and figure \ref{fig:taylor-error}.
The normalized errors are calculated as
\begin{align}
  \label{eq:error}
  E_{k} = \frac{\| \text{numerical solution}- \text{reference}\|_k}
    {\|\text{reference}\|_k} \ \text{,}
\end{align}
where the reference is given by expression \eqref{eq:OurTaylorS}, and where the subindex $k=1,\infty$ refers to one of two norms used,
\begin{alignat*}{2}
L_1 \ :  & \quad & \|\mathbf{Q}\|_1 &= \sum_i|Q_i| \ \text{,}\\
L_\infty \ : & & \|\mathbf{Q}\|_\infty &= \max_i|Q_i| \ \text{,}
\end{alignat*}
in this case evaluated over a 10-unit-long straight line normal to the propagation direction.
Convergence rates were calculated using the slope of a linear regression of $\log(E_k)$ as a function of $\log(\eta)$.
Note that the values of $E_{\infty}$ shown in table \ref{tab:error-ts} confirm the validity of the use of equation \eqref{TS:approx:E}:
since $\|\text{reference}\|_{\infty} = O(\epsilon) = 10^{-7}$ and our best value of $E_{\infty}$ is order $10^{-4}$, then the difference between analytic and numerical solutions is greater than $10^{-11}$, which is in turn greater than $O(\epsilon^2)=10^{-14}$.
It can also be observed that convergence rates are second order in the best cases, as is to be expected, and of an order greater than one in all cases. 
We believe these rates of convergence could possibly be improved through modification of the diffusive part of the numerical method. 
In results not reported here we have observed that convergence rates diminish for larger values of $\tilde{\delta}/\beta$, which is consistent with the fact that equation \eqref{TS:approx:E} is only valid for small amplitudes.
The behavior of the errors as functions of $\eta$ is monotonically decreasing as the mesh is refined ($\eta \rightarrow \infty$), reaching, when $\eta=82$, values below $0.3\%$ for $E_1$, and below $2\%$ for $E_\infty$.

The dependence of errors on $\nu_W$ is not the same for all cases: we have observed both increasing and decreasing behaviors, depending on the values of $\theta_T$ and $\eta$.
This is not surprising since smaller time steps imply more steps are needed to reach $t=100$. 
Indeed, it is known that the best choices for the CFL number depend on specific circumstances \citep{laney1998computational}. In all cases we observe this dependence is diminished when the mesh is refined.
In any case, $\nu_W\approx 1$ are to be preferred as they often lead to better results \citep{leveque-book-fv}, and time steps will certainly be larger, so that execution times will be smaller.
Tests for $\nu_W>1$ were attempted, but they yielded only unstable results.

Results in figure \ref{fig:taylor-error} also illustrates how errors diminish when the propagation is aligned with the mesh. 
This is also expected since in this case numerical fluxes point in the direction of physical fluxes. 
This dependence is reduced when the mesh is refined, and for the cases with $\eta=82$ the difference does not exceed $2\%$.

\subsection{Validation Against a Set of Full Wave Numerical Results}
\label{sec:agains-anoth-numer}

\begin{table*}
    \begin{tabular}{ccccccccc}
      \toprule
      & & \multicolumn{3}{c}{Convergence rate} &
      \multirow{2}{*}{\parbox{1.8cm}{\centering Worst $R^2$ value}} 
      & \multicolumn{2}{c}{Error on $\eta=82$} \\
      \cmidrule(r){3-5} \cmidrule(r){7-8}
      & Error & Best & Worst & Mean & & Best & Worst \\
\midrule
\multirow{2}{*}{$a=10$} &
$E_1$ &
2.9841 & 0.3887 & 2.0830 & 0.9663 & 0.0057 & 0.0241 \\
& $E_\infty$ &
2.6415 & 0.5946 & 2.0304 & 0.9653 & 0.0103 & 0.0269 \\
\midrule
\multirow{2}{*}{$a=20$} &
$E_1$ &
2.5039 & 1.0729 & 2.0515 & 0.9790 & 0.0075 & 0.0315 \\
& $E_\infty$ &
2.4766 & 1.3016 & 2.0568 & 0.9810 & 0.0109 & 0.0381 \\
\midrule
\multirow{2}{*}{$a=30$} &
$E_1$ &
2.3531 & 1.0381 & 1.9283 & 0.9868 & 0.0093 & 0.0361 \\
& $E_\infty$ &
2.2553 & 1.4884 & 1.9648 & 0.9787 & 0.0180 & 0.0338 \\
      \bottomrule
    \end{tabular}
  \caption{Focused wave, convergence analysis.}
  \label{tab:error-hifu}
\end{table*}

\begin{figure*}[htb]
\centering
\framebox{
  \resizebox{\textwidth}{!}{
    \includegraphics{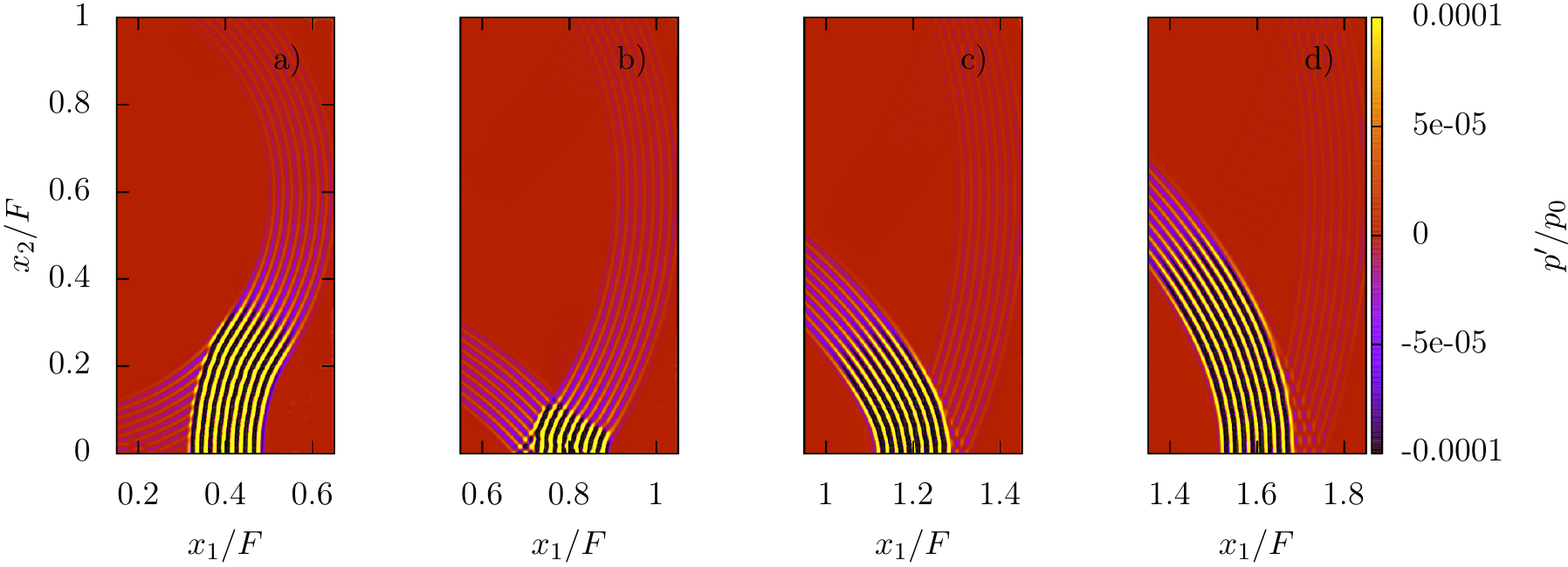}
  }
}
\caption{Focused wave, view from above of numerical results for $p'/p_0$,  with $a=30$, and a) $t=20$, b) $t=40$, c) $t=60$, d)  $t=80$.
Scale for $p'/p_0$ is compressed to make edge waves more visible%
.}
\label{fig:hifu-upper}
\end{figure*}
\begin{figure}[htb]
\centering
\framebox{
  \resizebox{3.3in}{!}{
    \includegraphics{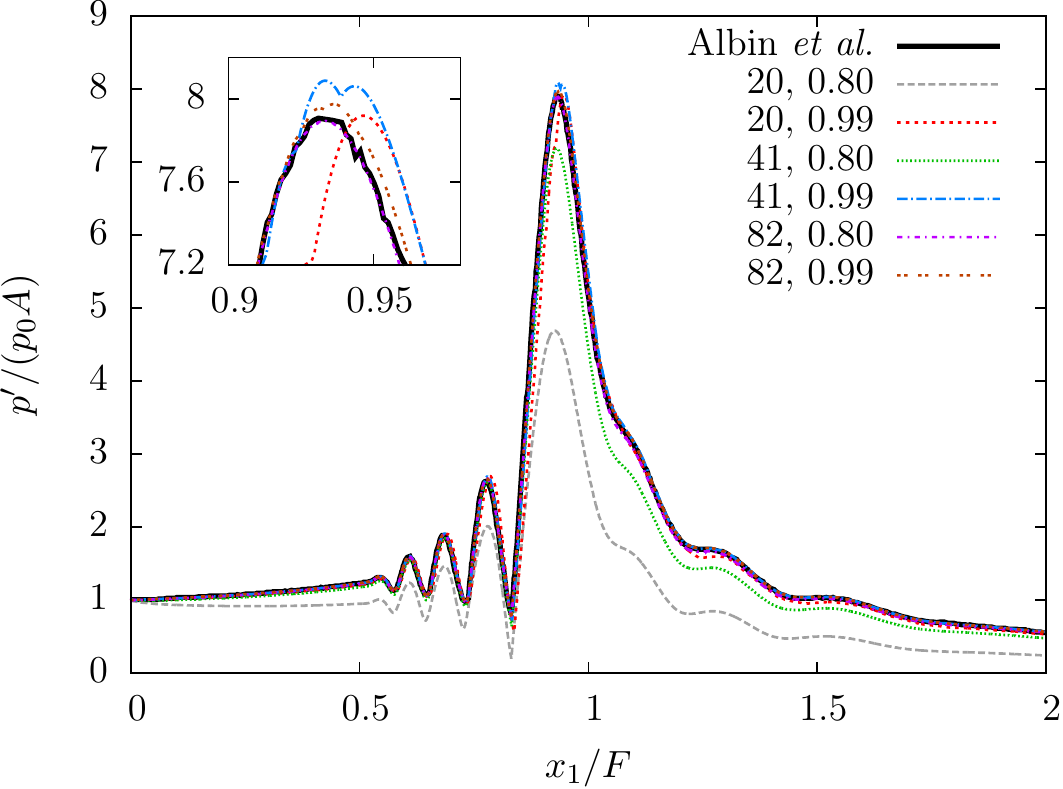}
  }
}
\caption{Focused wave, on-axis pressure maximum for $a=30$.
Numbers in labels correspond to values of $\eta$, $\nu_W$. 
Reference values taken from Albin {\em et al.} \citep{albin}, figure 10(c) DNS data series%
.}
\label{fig:hifu-u1max-a30}
\end{figure}
\begin{figure}[htb]
\centering
\framebox{
  \resizebox{3.3in}{!}{
    \includegraphics{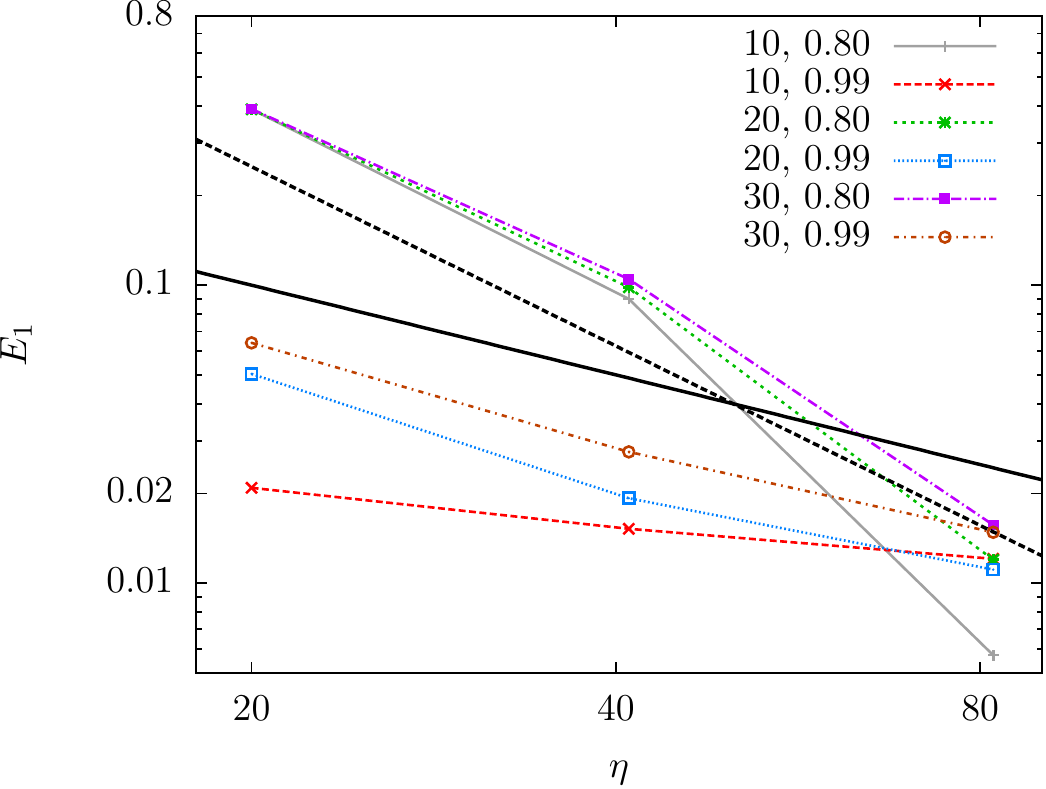}
  }
}
\caption{
Focused wave, error $E_1$ as a function of $\eta$.
Numbers in labels correspond to values of $a$, $\nu_W$.
For reference, rates of convergence as $\eta^{-1}$ and $\eta^{-2}$ are indicated by means of thick solid and thick dashed lines, respectively%
.}
\label{fig:hifu-error}
\end{figure}
\begin{figure}[htb]
\centering
\framebox{
  \resizebox{3.3in}{!}{
    \includegraphics{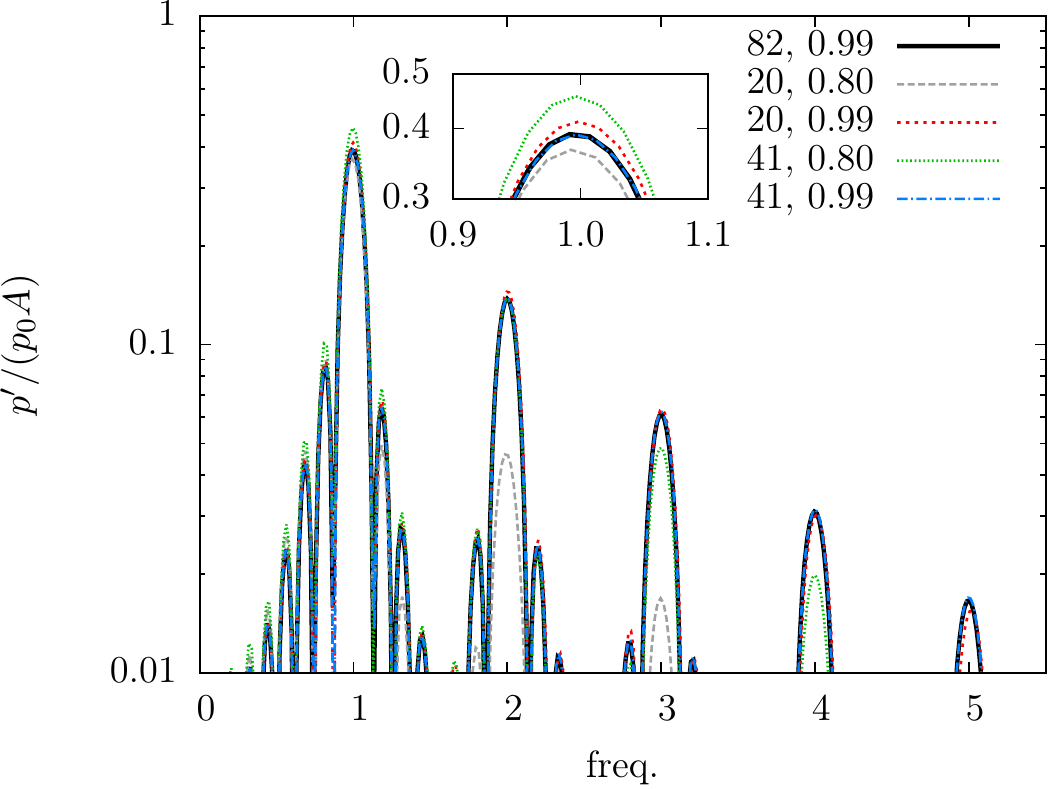}
  }
}
\caption{Focused wave, frequency spectrum, $a=30$.
Numbers in labels correspond to values of $\nu_W$, $\eta$%
.}
\label{fig:spec}
\end{figure}

For this second test the results of Albin {\em et al.} \citep{albin} were used as reference, and simulation conditions were reproduced as closely as possible.
The system studied in this case, viewed as a two-spatial-dimensional domain, is an approximation to the case where a transducer with the shape of a circle arc produces a focusing wave packet perturbation.
When viewed as a three-spatial-dimensional system, this corresponds to a transducer which results from a longitudinal cut of an infinitely long cylindrical shell, and not from a cut of a spherical shell.
The center of the aforementioned circle is located at the point $x_1=F$, $x_2=0$, with $F$ the focal length, and $R=\sqrt{F^2 + a^2}$ its radius.
Then the ends of the transducer are located at $x_1=0$, $x_2 = \pm a$.

The boundary conditions for this simulation vary according to the region:
for the first stage, and when $x_1=0$ and $0\leq~x_2~\leq~a$, the boundary conditions are obtained by means of
\begin{subequations}
  \label{eq:hifuBC}
  \begin{align}
    q_1(0,x_2,t) &= 1+g \ \text{,} \\
    q_2(0,x_2,t) &= \frac{F g}{\sqrt{x_2^2 + F^2}} \ \text{,} \\
    q_3(0,x_2,t) &= \frac{-x_2 g}{ \sqrt{x_2^2 + F^2} } \ \text{,} \\
    \label{eq:hifuBC:g}
    g(x_2,t) &=
    A  \sin\left(2\pi\tau \right)
    e^{ -\left( \frac{1}{4}\tau\right)^{-10}} \ \text{,}
  \end{align}
\end{subequations}
where $\tau = t + \frac{x_2^2}{2F}$.
This implies linear propagation is assumed in the small region between the transducer and $x_1=0$.
For the rest of the left border ($x_1=0$, $a < x_2$), and the top and right borders, a zero-order extrapolation has been used, which is a simplified version of absorbing boundaries \citep{leveque-book-fv}.
The bottom border, which is treated in a reflective manner, makes use of the symmetry of the problem and reduces the calculations by half.
In the second stage, the complete left border was taken to be a zero-order extrapolation and the other borders remained as they were.

In this case $L$, in equation \eqref{eq:renorm}, was taken to be equal to the wavelength, so that in dimensionless variables the wavelength is 1, and $\eta$ is the number of cells per wavelength.
The other parameters take values
\begin{align*}
  F = 50
  \ \text{,} \quad
  A = 4.217\times 10^{-4}
  \ \text{,} \\
  \tilde{\delta} = 2.974 \times 10^{-4}
  \ \text{,}  \quad
  \beta = 4.8
  \ \text{,}
\end{align*}
which have been nondimensionalized using values corresponding to soft tissue \citep{albin}: $\delta = 6.4117\times10^{-4}$, $c_0 = 1540 \, \text{m}/\text{s}$, $\rho_0=1000 \, \text{kg}/\text{m}^3$, and a driving frequency $f_0=1.1 \, \text{MHz}$.

Simulations were performed for the combinations of the following values: $a=10$, $20$, $30$; $\eta=20$, $41$, $82$; and $\nu_W=0.6$, $0.7$, $0.8$, $0.9$, $0.99$.
As pointed in out in the previous section, the $\nu_W$ value does not exactly match the real CFL value, which led us to measure standard deviation as well, and it was in all cases lower than $2.5\times10^{-4}$.
All the simulations were performed up until a time value $t=100$, that is, until the wave packet had traveled a distance 100 times larger than the wavelength.
A view from above of a typical simulation is shown in figure \ref{fig:hifu-upper}, where equations \eqref{eq:adim:q} and \eqref{eq:EOSlin} were used to recover physical variables.

The values reported by Albin {\em et al.} that we have used to make comparisons are the pressure maxima over the propagation axis, which correspond in our case to $x_2 = 0$.
In order to avoid a transitory state, the maximum pressure was measured using the central peak in the wave packet.

The comparison between schemes is illustrated, for pressure maxima over the propagation axis in figure \ref{fig:hifu-u1max-a30}.
The lobe structure, from $x_1=0.5$ to $x_1=0.8$, is a consequence of the interaction of the wave packet we introduce, and which can be identified in figure \ref{fig:hifu-upper} as the high contrast line pattern, with an edge wave formed at $x_1=0$, $x_2=\pm a$, because of diffraction, and which is visible in figure \ref{fig:hifu-upper} as a series of lighter line patterns.
The convergence analysis shown in table \ref{tab:error-hifu} and figure \ref{fig:hifu-error} was calculated as described in Section \ref{sec:agains-analyt-solut}, but in this case expression \eqref{eq:error} was evaluated for pressure maxima over the $x_1$-axis.
In this case, convergence rates are second-order, and both errors decrease monotonically as the mesh is refined, and are under $4\%$ for all cases with the finest mesh, $\eta=82$.
Conversely, a non-monotonic behavior of error as a function of $\nu_W$ is found, but, as we have pointed out at the end of the previous section, this is not a problem, and values of $\nu_W$ approaching $1$ are preferred.

The frequency spectrum of the signal is illustrated for the case $a=30$ in figure \ref{fig:spec}. It was calculated using a standard FFT over a time series taken at the point $x_1=1$, $x_2=0$.
See that because of the variable nondimensionalization in eq. \eqref{eq:renorm}, and the form of the signal, particularly eq. \eqref{eq:hifuBC:g}, the fundamental frequency in this case equals 1.
In this figure we can see that the cases with $\nu_W=0.99$ accurately describe the amplitudes of the first five harmonics, even when for this frequency the case $\eta=20$ provides only $4$ points per wavelength.
On the other hand, it is noticeable in figure \ref{fig:hifu-error} that for the case $\nu_W=0.99$, $\eta=20$, we obtain an error close to $10\%$, because of a slight shift, observable in figure \ref{fig:hifu-u1max-a30}.
For lower values of $\nu_W$, as illustrated in the same figure, such as $\nu_W=0.80$, the case $\eta=20$ has problems starting from the second harmonic, as does the case $\eta=41$ starting at the fourth harmonic, where both have $10$ points per wavelength. 
This suggests the case $\eta=82$ has an appropriate description of the amplitude up to the 8th harmonic.
Observations corresponding to higher harmonics are not reliable because they have amplitudes significantly smaller that the error we can estimate for our best case, of the order of $1\%$.

\subsection{Performance}
\label{sec:performance}

\begin{table}
  \begin{tabular}{llllc}
    \toprule
    & & $\eta$ & $\nu_W$ &
    \multicolumn{1}{c}{execution time} \\ 
    \midrule
    \multicolumn{1}{l}{
    \multirow{2}{*}{\parbox{75pt}{Taylor shock, \\
Section~\ref{sec:agains-analyt-solut}}}}
    & shortest & 5 & 0.99 &
    \multicolumn{1}{c}{4s} \\ 
    \cmidrule{2-5}
    \multicolumn{1}{c}{}
    & longest & 82 & 0.6  &
    \multicolumn{1}{c}{6min}  \\ 
    \midrule
    \multicolumn{1}{l}{
    \multirow{2}{*}{\parbox{75pt}{HIFU, \\
Section~\ref{sec:agains-anoth-numer}}}}
    & shortest & 20 & 0.99 &
    \multicolumn{1}{c}{55s} \\ 
    \cmidrule{2-5}
    \multicolumn{1}{c}{}
    & longest  & 82 & 0.6  &
    \multicolumn{1}{c}{47min} \\ 
    \bottomrule
    \end{tabular}
    \caption{Shortest and longest execution times.}
    \label{tab:times}
\end{table}

To make a performance comparison, a simulation was done with both a standard serial fortran CLAWPACK \citep{clawpack} 4.6.1 code and the C++/CUDA code described above.
A $256\times256$ mesh was used, with the first stage code only, and parameters
\begin{align*}
  \beta &= 4
  \ \text{,} \qquad
  \tilde{\delta} = 10^{-7}
  \ \text{,} \qquad
  L=\lambda
  \ \text{,} \qquad
  \text{$\eta$}=20
  \ \text{.}
\end{align*}
In the case where no data was written to disk and no plots were generated, the time execution factor, speedup, was of the order of $60$.
In practice, one needs to store and plot some partial results all the time, however.
Taking this into account, this factor can change depending on the plots wanted and the tool used to generate them, which is in both cases an external tool. 
However, in the case of GPU computation it is advantageous that direct visualization of the mesh is possible while the computation is being done, and this adds practically nothing to the execution time.
A different CUDA finite volume implementation, CUDACLAW \citep{cudaclaw}, has reported an execution time improvement of $30$ for CUDA execution over fortran execution.
The discrepancy with regards to our case could be due to the fact that different CPU processors were used, i7 in their case, and i3 in our case.
A more ambitious parallel finite volume project, ManyClaw \citep{terrel2012manyclaw}, currently in development state, implements more parallelization levels, including multiple CPU and GPU devices.

The shortest and longest execution times for the simulations presented in the previous sections are listed in Table~\ref{tab:times}. Even when these execution times are not too large, finer meshes could not be used because of memory limitations of the GPU.

With relation to the reference results \citep{albin} we have used in Section~\ref{sec:agains-anoth-numer}, authors report they have used a mesh with approximately $21$ points per wavelength, and CFL values of $1/40$ for the first stage, and $1/10$ for the second stage.
Performing a rough comparison with our method, using $\eta=82$, $\nu_W=0.99$, with a value of the errors under $2\%$, the total amount of spatiotemporal points for two spatial dimensions is only $1.5$ and $6$ times larger in our case, for the first and the second stage, respectively.
Albin {\em et al.} report for this case an execution time of $14$ min, on a 128 core cluster, which is approximately half the execution time in our case, $31$ min, where the code is run on a single machine.

In results not reported here, we have seen single precision implementations sometimes offer results as good as those obtained with double precision calculation.
The speedup in those cases for our code and hardware is approximately $1.5$, but it could be larger given a different hardware configuration.

\section{Discussion and Conclusions}
\label{sec:disc-concl}

A system of balance laws has been presented which constitutes a full wave model for nonlinear acoustic propagation at least as general as the Westervelt equation.
This system of balance laws is well suited for use with a finite volume method, which we have  implemented using the Roe linearization.
In the present work we have produced a two dimensional scheme, but the equations presented in Section~\ref{sec:equations} can be used directly for a three dimensional generalization.
In future stages of this work, the technical aspects of this implementation will be addressed.
Moreover, the set of equations \eqref{eq:OurSystem} could also be implemented using other time -- or frequency -- domain numerical methods.

Taking into account the restrictions of the model as listed in Section~\ref{sec:equations}, the presented method can still be applied to a variety of systems where nonlinear propagation is observed:
HIFU,
ultrasound imaging,
parametric acoustic arrays,
oblique shock interactions,
propagation in waveguides \citep{rendon2010nonlinear},
and underwater acoustics.

An important feature of our formulation is that the Lagrangian density has not been discarded at any point.
When this quantity is indeed negligible, we expect our system of conservation laws to be equivalent to Westervelt equation, but when it is not, as we imagine might be the case for a high-intensity focalized beam, we might obtain different results from models which do without this quantity altogether.
This exploration of the effect of the Lagrangian density will be subject of future work.

Recent applications using ultrasound imaging \citep{pinton2009heterogeneous, okita2011development} show that visualization is an important tool for understanding complex phenomena, and implementation on GPU systems allows for quick visualization of the acoustic field, linear or nonlinear.
In this way, GPU implementation of numerical methods, such as the one presented here, could constitute an important tool for understanding a variety of acoustic phenomena.
In this case, for example, the presence of a lobed structure in figure \ref{fig:hifu-u1max-a30} can visually be understood as the interference of waves present in figure \ref{fig:hifu-upper}.

The validation tests performed for the numerical method show a very good agreement with references, particularly with the finest mesh $\eta=82$, and large time steps $\nu_W=0.99$.
In the case of a plane wave analytic solution, the normalized error $E_1$ was under $0.3\%$, and in the case of full wave propagation it was under $1.5\%$.
Convergence rates were second order for the focused wave, and between first and second order for the Taylor shock case.
This behavior can be improved in future versions of the code by changing the numerical treatment of the diffusive part.
Comparisons among the numerical results presented here suggest our method uses 10 points per wavelength for the highest harmonic correctly simulated, which is up to 5 times larger than what is needed with other methods \citep{jing2012k}.
The performance of the scheme presented in this work compares favorably with the recent method we have used as reference \citep{albin}, execution times are comparable even taking into account the fact that our scheme is executed in a single machine equipped with a GPU, as opposed to a full-fledged cluster.
Some recent publications on frequency domain methods claim that one advantage of their approach is the execution time optimization, due to the larger point separation that can be achieved for its meshes \citep{albin, huijssen2010iterative}.
However, even in such cases, having different approaches to model the same system is advantageous because it permits the validation of numerical results, as is the case here.

This code is publicly available under a free and open-source software (FOSS) license in the address \url{https://github.com/rvelseg/FiVoNAGI/tree/1311.3004}.
The name of the repository stands for Finite Volume Nonlinear Acoustics GPU Implementation.
This version of the code is capable of reproducing the results reported in this paper.
As a precedent, there are other nonlinear acoustic codes published as FOSS \citep{frijlink2008abersim, anderson20002d}.

\section{Acknowledgments}
\label{sec:acknowledges}

The authors are grateful to Dr. Felipe Ordu\~{n}a, and Dr. Carlos M\'{a}laga for their useful comments,
to the CLAWPACK \citep{clawpack} project for sharing their base code,
to Dr. Randall LeVeque for the useful information provided,
and to Dr. Oscar P. Bruno, Dr. Nathan Albin, and coworkers for kindly providing comparison data.
This work was financially supported by PAPIIT IN109214, PAEP UNAM 2011-2014, and Conacyt M\'{e}xico.

\section*{References}

\bibliographystyle{unsrtnat}

\bibliography{references}

\end{document}